\documentclass[%
 reprint,
 prl,
 floatfix,
 amsmath,amssymb,
 aps, widetext, superscriptaddress
]{revtex4-2}

\usepackage{xcolor}
\usepackage{hyperref}
\usepackage{graphicx}
\usepackage{dcolumn}
\usepackage{bm}
\usepackage{microtype}
\usepackage{physics}
\usepackage{xspace}
\usepackage{chemformula}
\usepackage{csquotes}
\usepackage{lipsum}
\usepackage{comment}
\usepackage{todonotes}
\usepackage{color,soul}

\allowdisplaybreaks[3]

\newcommand{\tuwien}{Institute for Theoretical Physics, TU Wien, Vienna A-1040, Austria}
\newcommand{\wq}{\ensuremath{\omega_\mathrm{q}}\xspace}

\begin{document}
\title{Passive quantum state transfer in a dispersion-engineered waveguide}

\author{Zeyu Kuang}
\thanks{These authors contributed equally to this work.}
\author{Oliver Diekmann}
\thanks{These authors contributed equally to this work.}
\author{Lorenz Fischer}
\author{Stefan Rotter}
\author{Carlos Gonzalez-Ballestero}
\email{carlos.gonzalez-ballestero@tuwien.ac.at}

\affiliation{\tuwien}

\date{\today}

\begin{abstract}
High-fidelity state transfer is fundamentally limited by time-reversal symmetry: one qubit emits a photon with a certain temporal pulse shape, whereas a second qubit requires the time-reversed pulse shape to efficiently absorb this photon. 
This limit is often overcome by introducing active elements. Here, we propose an alternative solution: by tailoring the dispersion relation of a waveguide, the photon pulse emitted by one qubit is passively reshaped into its time-reversed counterpart, thus enabling perfect absorption. We analytically derive the optimal dispersion relations in the limit of small and large qubit-qubit separations, and numerically extend our results to arbitrary separations via multiparameter optimization.
We further propose a spatially inhomogeneous waveguide that renders the state transfer robust to variations in qubit separations. In all cases, we obtain near-unity transfer fidelity ($\geq 98\%$) that is robust against imperfections in parameter values and propagation loss. Our dispersion-engineered waveguide provides a compact and passive route toward on-chip quantum networks, highlighting dispersion as a powerful resource in waveguide quantum electrodynamics.
\end{abstract}

\maketitle

\begin{figure}[b!]
    \centering
    \includegraphics[]{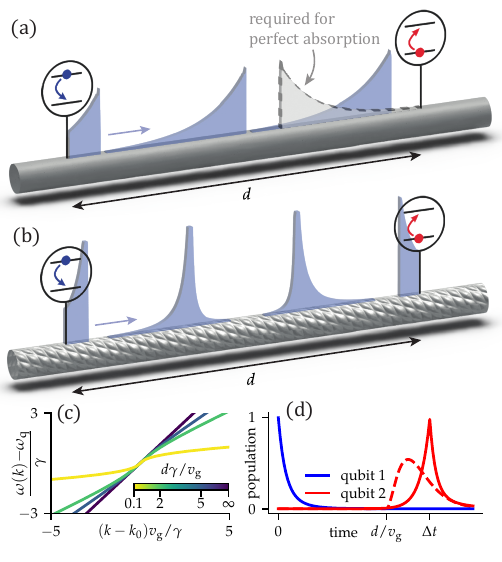}
    \caption{State transfer between two qubits coupled to a chiral waveguide with (a) linear and (b) engineered dispersions. Snapshots of the emitted photon pulse are shown at different times. (c) Dispersion relations required for perfect state transfer (see text for details) for different qubit-qubit distances $d$ (normalized by the pulse width $v_\mathrm{g} / \gamma$). (d) Qubit population dynamics in the  dispersion-engineered waveguide (solid lines) and the linear-dispersion waveguide (dashed red line for the second qubit).} 
    \label{fig:1}
\end{figure}

Quantum networks are key platforms for quantum communication~\cite{braunstein_teleportation_1998,furusawa_unconditional_1998,zheng_persistent_2013}, distributed computation~\cite{chang_single-photon_2007, zheng_waveguide-qed-based_2013,mahmoodian_quantum_2016}, and fundamental studies of many-body physics~\cite{chang_crystallization_2008,zheng_waveguide_2010,mahmoodian_dynamics_2020,armon_photon-mediated_2021,tabares_variational_2023}. They rely on the faithful transfer of quantum states between distant nodes~\cite{cirac_quantum_1997,kimble_quantum_2008,northup_quantum_2014,blais_circuit_2021}. 
Waveguide quantum electrodynamics (QED) provides a particularly natural implementation~\cite{sheremet_waveguide_2023,gonzalez-tudela_lightmatter_2024,lodahl_chiral_2017}, as waveguides can efficiently channel photons between spatially separated qubits. Yet even in the ideal case of two identical qubits coupled to a perfectly chiral waveguide~\cite{lodahl_chiral_2017, de_bernardis_light-matter_2021, de_bernardis_chiral_2023}, where every photon emitted by the first qubit is guided toward the second [Fig.~\ref{fig:1}(a)], perfect transfer is limited by a fundamental mismatch: the first qubit emits a photon with a certain temporal pulse shape, whereas the second qubit requires a time-reversed pulse shape to absorb the photon efficiently~\cite{stobinska_perfect_2009, leuchs_time-reversal_2012, aljunid_excitation_2013}.
In the Markovian limit, 
this pulse asymmetry limits the excited-state occupation of the second qubit to a maximum of $4/e^2 \approx 0.54$~\cite{stobinska_perfect_2009}, regardless of the group velocity, distance, or qubit-waveguide coupling rate.

A wide range of solutions have been proposed to overcome this state transfer limitation. The seminal work of Cirac \textit{et al.}~\cite{cirac_quantum_1997} introduces an external laser to dynamically tune the effective qubit-waveguide coupling rates, thereby forcing the emission of a time-reversal-symmetric pulse. This idea has been demonstrated experimentally~\cite{ritter_elementary_2012,reiserer_cavity-based_2015,northup_quantum_2014} and extended theoretically to account for pulse distortions~\cite{penas_improving_2023}, dissimilar atoms~\cite{randles_quantum_2023}, and multiplexed state transfer~\cite{penas_multiplexed_2024}. The principle of this active control can be generalized by replacing the external driving laser by other active elements such as a tunable cavity~\cite{thyrrestrup_non-exponential_2013, penas_improving_2023}, an intermediate atom with tunable coupling rate~\cite{li_flying-qubit_2022}, or an atom ensemble~\cite{hush_quantum_2016}.
A potential alternative route, which remains unexplored for quantum state transfer, is to modulate the waveguide to time-reverse the propagating pulse via a time-lens~\cite{donohue_spectrally_2016,karpinski_bandwidth_2017,raymer_time_2018}. The general principle behind these diverse approaches is that they all use active, time-dependent modulation to time-reverse the pulse and therefore achieve perfect state transfer. While versatile, these active components require additional monitoring and control,  making the overall system more complex.

In this Letter, we 
propose a passive method for perfect quantum state transfer, namely to  engineer the \textit{dispersion relation} of the waveguide. The resulting dispersion-engineered waveguide time-reverses the emitted pulse [see Fig.~\ref{fig:1}(b)] by providing the necessary phase shift at each frequency.
Based solely on linear and passive elements, our proposal can be potentially realized using standard microwave and photonic engineering techniques~\cite{lodahl_interfacing_2015,gu_microwave_2017,tianStaticHybridQuantum2021, gonzalez-tudela_lightmatter_2024}. It suggests new opportunities of photonic design in quantum optics~\cite{chang_quantum_2006, gonzalez-tudela_entanglement_2011, gonzalez-tudela_subwavelength_2015} and highlights engineered, dispersive few-photon pulse propagation as a potential powerful asset in waveguide QED.

\begin{figure*}[!t]
    \centering
    \includegraphics[]{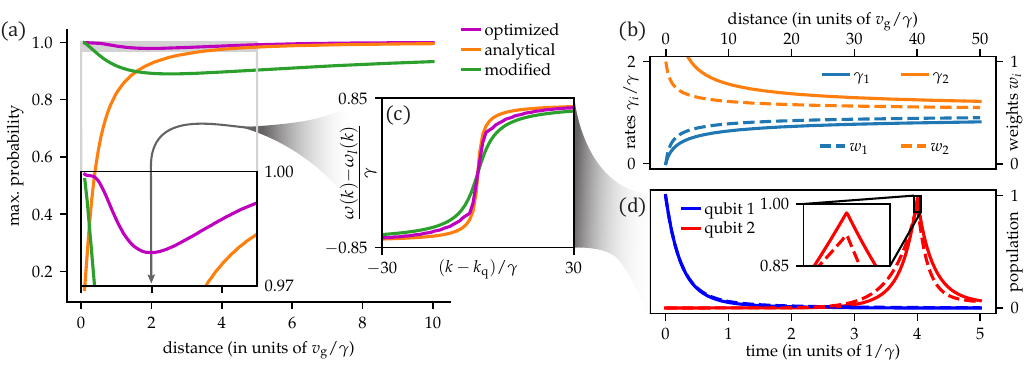}
    \caption{(a) Maximum absorption probability of the second qubit using the analytical [orange, see Eq.\eqref{eq:dispersion}], small-distance [green, see Eq.~\eqref{eq:dispersion_d0}] and numerically optimized [purple] dispersion relations. As before, we assume $\gamma=\pi\times10^{-4}\omega_\mathrm{q}$. In the small-distance limit, the first qubit decays biexponentially under the analytical dispersion of Eq.~(\ref{eq:dispersion}). The decay rates and weights [defined in Eq.~\eqref{eq:biexponential}] are plotted against the qubit distance $d$ in (b). (c) Dispersion relation at $d=2v_\mathrm{g}/\gamma$ where the linear contribution $\omega_l \equiv \omega\Delta t/d$ has been subtracted for better visualization. (d) Populations as a function of time for the dispersion relations of panel (c). The dashed lines are for the analytical dispersion; the solid lines are for the numerically optimized dispersion.}
    \label{fig:2}
\end{figure*}

We consider the system shown in Fig.~\ref{fig:1}(a): two identical qubits, located at positions \(x_1 = 0\) and \(x_2 = d\), chirally coupled to a single-band, one-dimensional electromagnetic waveguide. In the rotating-wave approximation, the Hamiltonian reads (\(\hbar = 1\))
\begin{equation}
\begin{aligned}
H &= \sum_{i=1}^{2}\omega_\mathrm{q} \sigma_i^+\sigma_i^- 
   + \int_{0}^{\infty}\!dk\,\omega(k)a_k^\dagger a^{\phantom{\dagger}}_k  \\
  &\quad + \sum_{i=1}^{2}\int_{0}^{\infty}\!dk\,\big[e^{ikx_i}g(k)a_k\sigma_i^+ + \text{H.c.}\big],
\end{aligned}
\label{eq:H}
\end{equation}
where \(\omega_\mathrm{q}\) is the qubit transition frequency, \(a_k\) is the annihilation operator for right-propagating photons of wave number \(k\) and frequency \(\omega(k)\), and \(\sigma_i^\pm\) are the raising and lowering operators of qubit~\(i\). 
We assume the coupling is perfectly chiral~\cite{pichler_quantum_2015, gonzalez-ballestero_chiral_2015,maffei_directional_2024} (i.e., the qubits only couple to right-propagating photons) and spectrally flat around $\omega_\mathrm{q}$, such that we can approximate \(g(k) = g\,\theta(k)\) with the Heaviside function \(\theta(k)\). 
The system wavefunction in the single-excitation subspace is
\begin{equation}
|\psi(t)\rangle = \Big(\sum_{i=1}^{2} c_i(t)\sigma_i^{+} + \!\int_{0}^{\infty}\!dk\,c(k;t)a^{\dagger}_k\Big)|\mathrm{g}_1 \mathrm{g}_2,0\rangle,   
\label{eq:psi}
\end{equation}
where \(|\mathrm{g}_i\rangle\) denotes the ground state of qubit~\(i\), \(|0\rangle\) the vacuum of the waveguide field, and \(c_i(t)\) and \(c(k;t)\) the corresponding  amplitudes. We aim at maximizing $\vert c_2(t)\vert$ for an initial state where the first qubit is excited at $t=0$ while the second qubit and the field are in their ground states, i.e., \(c_1(0)=1\) and \(c_2(0)=c(k;0)=0\).

Our hypothesis that a nonlinear dispersion $\omega(k)$ can enable perfect quantum-state transfer is most transparent in the perturbative limit, where deviations from a linear dispersion are small.
In this limit, we can use the Markovian Wigner–Weisskopf theory of spontaneous emission~\cite{weisskopf_berechnung_1930, weisskopf_uber_1930} to represent the photon emitted by the first qubit at time $t=0$ classically as an electric field with an exponentially decaying envelope: $E(x_1,t) = \theta(t)e^{-(\gamma + i\omega_\mathrm{q})t}$ where the decay rate $\gamma = \pi g^2 / v_\mathrm{g}$ is governed by the coupling strength $g$ and on-resonance group velocity $v_\mathrm{g} = (d\omega/dk)|_{\omega=\omega_\mathrm{q}}$. Its frequency spectrum is $E(x_1,\omega) = 1/[\gamma - i(\omega - \omega_\mathrm{q})]$.
After propagating through a waveguide with an arbitrary single-band dispersion $k(\omega)$, the field at the position of the second qubit becomes $E(x_2,\omega) = E(x_1,\omega)e^{ik(\omega)d}$.
For perfect absorption at some time $t = \Delta t$, the second qubit must receive a field whose envelope is the \textit{time-reverse} of the emitted field~\cite{stobinska_perfect_2009}, i.e., $E_{\text{target}}(x_2,t) = \theta(\Delta t - t)e^{(\gamma - i\omega_\mathrm{q})(t - \Delta t)}$ or, in frequency domain, $E_{\text{target}}(x_2,\omega) = e^{i\omega \Delta t}/[\gamma + i(\omega - \omega_\mathrm{q})]$.
Equating $E(x_2,\omega) = E_{\text{target}}(x_2,\omega)$, we obtain the required the dispersion relation
\begin{equation}
    k(\omega) = \frac{\Delta t}{d}\,\omega - \frac{2}{d}\,\arctan\!\left(\frac{\omega - \wq}{\gamma}\right).
    \label{eq:dispersion}
\end{equation}
The transfer time $\Delta t$ can be determined by combining Eq.~(\ref{eq:dispersion}) with the definitions of $v_\mathrm{g}$ and $\gamma$, yielding  $\Delta t = d/v_\mathrm{g} + 2/\gamma$.
Equation~(\ref{eq:dispersion}) is the key dispersion that converts an exponentially decaying pulse at the position of the first qubit into its time-reversed, exponentially rising counterpart at the position of the second. This conversion process is illustrated in Fig.~\ref{fig:1}(b).

Figure~\ref{fig:1}(c) shows the dispersion of Eq.~(\ref{eq:dispersion}) for several qubit-qubit separations $d$ (here we take $\gamma=\pi\times10^{-4}\omega_\mathrm{q}$). Note that the group velocity $d\omega/dk$ peaks near the resonance frequency $\omega_\mathrm{q}$, indicating that near-resonant components travel faster than off-resonant ones. This behavior is precisely what the time-reversal conversion in Fig.~\ref{fig:1}(b) requires: to transform a decaying pulse into a rising one, the slowly varying tail of the exponential envelope (dominated by near-resonant frequencies) must overtake its rapidly varying front (where off-resonant frequencies reside). As the separation $d$ increases, the required difference in group velocity decreases, since the tail has more time to overtake the front; consequently, the dispersion becomes increasingly linear, as seen in Fig.~\ref{fig:1}(c). In this regime the dispersion is consistent with our perturbative assumption, implying that for sufficiently large $d$, the analytical dispersion of Eq.~(\ref{eq:dispersion}) should result in perfect state transfer.
This is demonstrated in Fig.~\ref{fig:1}(d) where
 already for $d = 5v_\mathrm{g}/\gamma$ (five pulse widths), the dispersion of Eq.~(\ref{eq:dispersion}) results in a 98\% excitation probability of the second qubit (solid line) at the transfer time $\Delta t$, far above the 54\% maximum attainable with a linear dispersion (dashed line). At larger distances the excitation probability asymptotically approaches 100\%. These excitation probabilities are obtained by exact diagonalization of the Hamiltonian in Eq.~(\ref{eq:H}) within the single-excitation subspace.

Although the dispersion of Eq.~(\ref{eq:dispersion}) enables near-perfect state transfer for qubits that are far apart, it fails when the qubits are brought close together. This can be seen in Fig.~\ref{fig:2}(a) where the maximum excitation probability of the second qubit (orange curve), calculated using the dispersion of Eq.~(\ref{eq:dispersion}), drops rapidly as the qubit distance $d$ decreases. The origin is evident in Fig.~\ref{fig:1}(c): when the distance $d$ becomes comparable to $v_\mathrm{g}/\gamma$ (the pulse width), the required dispersion becomes highly nonlinear. In this regime, the Markovian approximation no longer holds, and the qubit no longer decays exponentially, so that its emitted photon no longer has an exponential pulse envelope. Consequently, the phase compensation prescribed by Eq.~(\ref{eq:dispersion}) becomes inaccurate, and perfect state transfer is lost.

To overcome this challenge in the small-$d$ regime, we refine our previously derived dispersion through an iterative procedure. In our previous derivation, we have assumed a nearly linear dispersion to obtain an exponential emission profile, and used it to derive the dispersion of Eq.~(\ref{eq:dispersion}). We now take this dispersion as input to \textit{update} the emission profile and, from it, construct a refined dispersion. As the new ``input dispersion'' of Eq.~(\ref{eq:dispersion}) is now nonlinear, Wigner--Weisskopf theory cannot be applied. We instead analytically and exactly compute the full qubit dynamics using the resolvent method (see End Matter). The resulting excitation amplitude of the first qubit is
\begin{equation}
    c_1(t) = w_1 e^{-(i\omega_\mathrm{q}+\gamma_1)t} - w_2 e^{-(i\omega_\mathrm{q}+\gamma_2)t},
    \label{eq:biexponential}
\end{equation}
where $\gamma_1 = \xi - \sqrt{\xi^2 - \gamma^2}$, $\gamma_2 = \xi + \sqrt{\xi^2 - \gamma^2}$, $\xi = \gamma + v_\mathrm{g}/d$, and $w_i = (\gamma_i - \gamma)/(\gamma_1 - \gamma_2)$. Its decay is thus no longer exponential but \textit{bi}-exponential. Figure~\ref{fig:2}(b) shows how the two decay rates $\gamma_{1,2}$ and their weights $w_{1,2}$ depend on the qubit distance $d$: for $d \gg v_\mathrm{g} / \gamma$, both decay rates approach the Markovian value: $\gamma_{1,2} \rightarrow \gamma$, recovering the single-exponential decay. For smaller $d$, the two rates become different, explaining why the dispersion of Eq.~\eqref{eq:dispersion} fails to yield perfect transfer for small $d$: the dispersion is designed to invert a single exponential pulse, yet the emitted pulse under this dispersion is biexponential and therefore not perfectly inverted.
We continue our iterative procedure by computing the photon amplitudes $c_k(t)$ analytically, and use them to extract a new optimal dispersion relation $\omega(k)$. Unfortunately, we find there is no single-band solution for the dispersion relation (see End Matter). However, we can harness the fact that, as evidenced by Fig.~\ref{fig:2}(b), for sufficiently small separations ($d \ll v_\mathrm{g}/\gamma$), the biexponential weights tend to $w_1 \to 0$ and $w_2 \to -1$, so that the emission of qubit 1 becomes again dominated by a single exponential with modified decay rate $\gamma_2$. 
This indicates that in this limit, the dispersion of Eq.~(\ref{eq:dispersion}), which is designed to optimally invert an exponentially decaying pulse, should also lead to perfect absorption under the substitution of the Markovian decay rate $\gamma$ by the modified rate $\gamma_2$, i.e.,  
\begin{equation}
    k_{d \rightarrow 0}(\omega) = \frac{\Delta t}{d}\,\omega - \frac{2}{d}\,\arctan\!\left(\frac{\omega - \wq}{\gamma_2}\right).
    \label{eq:dispersion_d0}
\end{equation}
The maximum qubit occupation corresponding to this dispersion is shown by the green curve in Fig.~\ref{fig:2}(a). As expected, this  dispersion successfully restores perfect quantum state transfer in the $d \ll v_\mathrm{g}/\gamma$ limit, precisely where the original dispersion of Eq.~(\ref{eq:dispersion}) fails (orange curve). 
Remarkably, by choosing one of our two analytical expressions in the large-$d$ and small-$d$ limit, state transfer with probability $>90\%$ can be achieved for any separation $d$.

\begin{figure*}[tb]
    \centering
    \includegraphics[]{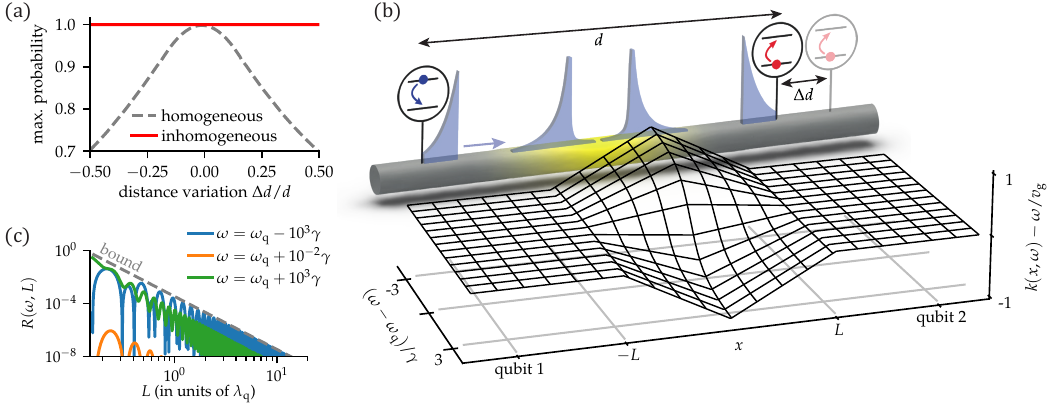}
    \caption{(a) Maximal excitation probability of the second qubit (dashed gray line) versus deviation $\Delta d$ from the assumed separation $d$ [see (b) for an illustration]. The state transfer fidelity decreases with $|\Delta d|$. We propose to solve this by a spatially inhomogeneous waveguide, whose dispersion (minus the linear part, for visibility) is shown by the 3D mesh grid in (b). This dispersion varies adiabatically in space within a length $2L$, minimizing reflection and effectively suppressing it for $L\gtrsim \lambda_\mathrm{q}\equiv2\pi v_\mathrm{g}/\omega_\mathrm{q}$ (panel c).  The dispersion of this segment is chosen to time-reverse the pulse. As a result, the maximum  occupation of qubit $2$ is near unity regardless of the distance variation [(a), red line].}   
    
    \label{fig:3}
\end{figure*}

To increase the fidelity beyond that achieved by the analytical dispersions of Eqs.~\eqref{eq:dispersion} and~\eqref{eq:dispersion_d0}, we perform a systematic multiparameter numerical optimization of the dispersion relation. For each qubit separation $d$, we initialize the optimization with the analytical dispersion of Eq.~(\ref{eq:dispersion}) and allow its shape to vary freely in order to maximize the excitation probability of the second qubit,
$|c_2(\Delta t, \omega(k))|^2$.
Both the dispersion curve $\omega(k)$ and the transfer time $\Delta t$ (this time is not known a priori) are our optimization variables. We parameterize $\omega(k)$ on a finite frequency grid (roughly 2000 points) centered around $\omega_\mathrm{q}$ and evaluate the resulting two-qubit dynamics by exact diagonalization at each iteration. The gradient of the objective function $|c_2(\Delta t, \omega(k))|^2$ with respect to $\omega(k)$ and $t$ is obtained via the adjoint method~\cite{errico_what_1997, cao_adjoint_2003, johnson2012notes}.  
The performance of the obtained optimized dispersion  is shown by a purple curve in Fig.~\ref{fig:2}(a), and recovers the analytical limits at large and small $d$. This dispersion markedly improves transfer fidelity in the intermediate separation regime; the largest gains occur near $d \approx 2v_\mathrm{g}/\gamma$, where the optimized dispersion [purple line in Fig.~\ref{fig:2}(c)] interpolates between the two analytically derived dispersions in the large- and small-separation limits, and increases the second-qubit excitation probability from 92\% to 98\% [Fig.~\ref{fig:2}(d)]. Note that the optimized dispersion does still not achieve unit transfer fidelity, with a residual infidelity that becomes maximal ($\sim 2\%$) at $d = 2v_\mathrm{g}/\gamma$.
This may stem from our restricted parameter search [currently restricted to dispersions close to Eq.~(\ref{eq:dispersion})] or it may reflect a limitation of single-band dispersion engineering.
On the other hand, the qubit dynamics is invariant under the transformation  $d \!\to\! d/s,\; g \!\to\! g/\sqrt{s},\; \omega(k) \!\to\! \omega(k/s)$ for any $s > 0$. This means that 
one can take as a reference a configuration $\{d,g,\omega(k)\}$ where perfect transfer is possible, and then fabricate a system where one engineers a coupling strength $g/\sqrt{s}$ and dispersion $\omega(k/s)$. This would enable the perfect transfer at another distance $d/s$. This is one way to reduce the previous 2\% residual infidelity in the intermediate separation regime. Another way is to use a spatially-varying waveguide which we present in the following.


So far, all our proposed dispersions (either analytically derived or computationally optimized) are optimized for a given fixed  qubit-qubit separation
$d$. As a consequence, their performance for state transfer degrades when qubit $2$ is imperfectly positioned at a distance $d+\Delta d$ [illustrated in Fig.~\ref{fig:3}(b)]. This is shown by the dashed gray line in  Fig.~\ref{fig:3}(a), which corresponds to the occupation of qubit $2$ as a function of $\Delta d$, for $d = 15v_\mathrm{g}/\gamma$ and $\gamma=\pi\times10^{-4}\omega_\mathrm{q}$. Although the state transfer is still better than in linear-dispersion waveguides, the fidelity is far from $100\%$ for significant deviations $\Delta d$, decreasing below $\sim$90\% for $\Delta d=d/4$ and below 70\% for $\Delta d=d/2$. 
We find very similar values for the deterioration of fidelity for other values of $d$. 
The physical reason behind this deterioration is that the dispersion relation is chosen to perfectly time-reverse the pulse at a distance $d$ from qubit $1$. Since the dispersion is nonlinear, propagation by an extra distance $\Delta d$ distorts the pulse into a shape that is not optimally absorbed by the second qubit. 
The further the extra distance $\Delta d$, the more the pulse shape will distort, and the worse the excitation of qubit $2$ will be.

We solve this problem by placing the two qubits on two \textit{dispersionless} sections of the waveguide and only engineer the dispersion in between. This new architecture is illustrated in Fig.~\ref{fig:3}(b) with the dispersionless region marked by gray and the dispersive region by yellow. 
The two qubits can be placed anywhere in the dispersionless region, and the pulse emitted by the first qubit will always decay exponentially and propagate toward the dispersive region without distortion. The dispersive region occupies a length $2L$ and is dispersion-engineered to provide a frequency-dependent phase shift $\Delta\phi(\omega)=-2\arctan[(\omega-\wq)/\gamma]$, which converts the exponentially decaying pulse to an exponentially rising one [see Eq.~(\ref{eq:dispersion})]. More specifically, this phase shift is implemented through a spatially inhomogeneous dispersion profile, $k(\omega, x) = \omega / v_\mathrm{g} + W(x)\Delta\phi(\omega)(1-|x|/L)/L$, where $v_\mathrm{g}$ denotes the group velocity in the dispersionless regions, and $W(x) = \theta(x+L)-\theta(x-L)$ is a window function with $\theta(x)$ the Heaviside function.  
The nonlinear part of this dispersion, illustrated in Fig.~\ref{fig:3}(b), gives exactly the required total phase shift: $\int_{-L}^{L}[k(x,\omega) - \omega/v_\mathrm{g}]\,dx = \Delta\phi(\omega)$. The spatial variation of the dispersion must be adiabatic to avoid unwanted reflections. Our analytical calculation (see End Matter) shows that in the large-$L$ limit, the reflection is bounded by $R(\omega) \leq [\lambda/(2\pi L)]^4$, where $\lambda = 2\pi v_\mathrm{g} / \omega$ is the wavelength of light in the dispersionless region. This is shown in Fig.~\ref{fig:3}(c), where both on- and off-resonant frequencies decay rapidly for large $L$ according to the bound.  The resulting inhomogeneous waveguide can be described by a Hamiltonian similar to Eq.~(\ref{eq:H}), and its dynamics is solved by exact diagonalization (see End Matter). As shown by the red line in Fig.~\ref{fig:3}(a), with $L = 3 v_\mathrm{g} / \gamma$, the excitation probability of the second qubit reaches perfect fidelity for all qubit separations, in stark contrast to the homogeneous dispersion case, which is optimal for only one specific distance.
Our inhomogeneous design shows that by leveraging the additional spatial degrees of freedom in the dispersion, we can provide full robustness of state-transfer against distance variation of the qubits.

Last, we discuss potential experimental platforms that could implement our dispersion-based quantum state transfer. 
A key parameter is the spectral resolution required by our proposed dispersion, which is set by the emission rate $\gamma$ of the qubit [see Eq.~(\ref{eq:dispersion})]. 
Since usually $\gamma \ll \omega_\mathrm{q}$, our protocol requires engineering sharp spectral features in the dispersion. 
This could be realized in microwave superconducting circuits, where alternative routes for state transfer are actively sought~\cite{xuCoherentPopulationTransfer2016,yangQuantumInformationTransfer2004}, and whose dispersion can be engineered with frequency resolution equal to the qubit's emission rate (around $10\text{--}100$ MHz~\cite{astafiev_resonance_2010,van_loo_photon-mediated_2013,mirhosseini_cavity_2019}) using slow-light waveguides~\cite{othman_experimental_2017,mirhosseiniSuperconductingMetamaterialsWaveguide2018,zheng_degenerate_2020,ferreira_collapse_2021}. 
Another notable example in the microwave domain are surface acoustic waves, whose coupling to a superconducting qubit can result in emission rates as high as $\sim$100 MHz~\cite{bienfait_phonon-mediated_2019}, and whose dispersion can be engineered down to $\sim 10$MHz~\cite{shao_phononic_2019,bijay_1d_2019}. 
Although more challenging, our method could also be realized in the optical domain. For instance, using tailored cavities~\cite{zhangStronglyCavityEnhancedSpontaneous2018, liu_high_2018}, the emission rate in the optical regime can reach $5\text{--}44$ GHz, a spectral resolution can potentially be achieved by slow-light optical waveguides such as coupled optical resonators~\cite{poonTransmissionGroupDelay2006,pernice_high_2012,wu_micro-ring_2017}.   
All-pass filter designs~\cite{cohnDirectCoupledResonatorFilters1957, ferreira_collapse_2021} offer a systematic route to engineer such sharp dispersion without introducing additional loss.
Note that these two features (chiral emission enhancement and dispersion engineering) can be designed separately in different regions of a waveguide, see example in Fig.~\ref{fig:3}(b).


In summary, we propose a route to perfect single-photon state transfer in waveguide QED by engineering the dispersion of the wave\-guide. We analytically derive the required dispersion for qubits that are far apart [Eq.~\eqref{eq:dispersion}] and close-together [Eq.~\eqref{eq:dispersion_d0}], and numerically optimize it for regions in between. We also propose a new waveguide architecture that preserves perfect transfer even when the qubit positions are not correctly determined or later reconfigured. 
Besides the distance deviation, we show in End Matter that our system is robust in the presence of typical photon propagation loss and additional qubit decay channels. It is also robust against imperfections in qubit frequencies, qubit-waveguide couplings, and deviations from the optimal dispersion relation.
Our approach is fully passive during propagation and may be used for highly efficient photon detection~\cite{lescanne_irreversible_2020,lednev_spatially_2025} and quantum state distribution~\cite{kendon_perfect_2011,meignant_distributing_2019}. 
It can potentially reduce the complex time modulation required in the current quantum-memory protocols~\cite{axlineOndemandQuantumState2018} to a simple binary switch~\cite{bialczak_fast_2011,hangleiter_robustly_2024} that turns off the coupling after the quantum state transfer completes.
Our approach could also be generalized to the transfer of multi-excitation states between emitter ensembles~\cite{shen_strongly_2007,shen_strongly_2007-1,srivathsan_reversing_2015, mahmoodian_dynamics_2020}, such as entangled~\cite{kraus_discrete_2004, agusti_autonomous_2023} or collective superradiant states~\cite{asenjo-garcia_exponential_2017, pennetta_collective_2022, cardenas-lopez_many-body_2023}. It applies not only to photonic waves, but also to spin waves, acoustic waves, and matter waves, as dispersion is a fundamental concept across all these platforms.

\begin{acknowledgments}
This research was funded in part by the Austrian Science Fund (FWF) [10.55776/PAT1177623], [10.55776/COE1] and the European Union – NextGenerationEU.
The computational results have been achieved using the Austrian Scientific Computing (ASC) infrastructure. We thank Carmen Buchegger for useful discussions. 
\end{acknowledgments}

\bibliography{bib.bib}

\clearpage
\section{End Matter}

\subsection{Resolvent solution to nonlinear dispersion}
\label{sec:appendix:resolvent}
The dynamics of a single qubit coupled to a dispersion-engineered waveguide can be solved using the resolvent method~\cite{cohen-tannoudji_atom-photon_2008, gonzalez-tudela_markovian_2017}. To this end, we calculate the self-energy,
\begin{align}
    \Sigma(\delta+i0^+) &= g^2\int d{k}\frac{1}{{\delta}-\delta(k)+i0^+}\notag\\
    &=-\frac{i\pi g^2}{d}\left[ {\Delta t} +\frac{2}{{i\delta}-\gamma}\right],
\end{align}
where $\Delta t$ is the time until absorption, $\delta\equiv\omega-\wq$ and $\delta(k)\equiv\omega(k)-\omega_\mathrm{q}$. To derive the above expression, we substitute $d{\delta} = k'(\delta)\, d{k}$ and, exploiting that $k'(\delta)$ is meromorphic [see Eq.~\eqref{eq:dispersion}], employ the Sokhotski–Plemelj and residue theorem. We use this self-energy to describe the exact dynamics of the initially excited first qubit, 
\begin{equation}
    c_1(t) = e^{-i\omega_\mathrm{q}t}\frac{i}{2\pi}\int_{-\infty}^{\infty} d{\delta}\frac{e^{-i\delta t}}{\delta-\Sigma(\delta+i0^+)},
\end{equation}
which can again be solved using the residue theorem. For the integrand we find poles at $\delta_0 = i\gamma$, $\delta_{1,2}=-i\gamma_{1,2}$, which yields the expressions of Eq.~\eqref{eq:biexponential} in the main text upon closing the contour in the lower half plane. 

From the qubit dynamics in the nonlinear dispersion, we can derive the solution for the emitted pulse,
\begin{equation}
    c(k;t) =-g\sum_{i=1}^2 w_i\frac{e^{-(\gamma_i+i\wq)t}-e^{-i(\delta(k)+\wq)t}}{\delta(k)+i\gamma_i}\,.
\end{equation}
Fourier transforming in $k$ and $t$ gives
\begin{align}
    c(x_1,\delta) =  g\sum_{i=1}^2\frac{w_i}{\delta+i\gamma_i} \left[\frac{1}{v_\mathrm{g}} + \frac{2}{d\gamma} - \frac{i}{d(\delta+i\gamma)} \right]\,.
    \label{eq:pulseCorrected}
\end{align}
We can now use this expression to derive a corrected dispersion relation in analogy to the derivation in the main text.
Equation~\eqref{eq:dispersion} can be rewritten for arbitrary pulse shapes,
$    k(\omega) = \frac{\Delta t}{d}\,\omega - \frac{2}{d}\,\arg\!\left[c(x_1,\delta)\right].$
Inserting Eq.~\eqref{eq:pulseCorrected}, we find that the resulting $k(\omega)$ is not uniquely invertible, i.e., it cannot be implemented using a single band for $d\lesssim 1.7 v_\mathrm{g}/\gamma$.
While a multi-band dispersion is not fundamentally impossible to engineer, our iterative process, which assumes a single band, halts here.

\subsection{Solution of a linear ramp}
In this section, we derive the transmission and reflection coefficients of the linear ramp in Fig.~\ref{fig:3}(b). For convenience, we separate the dispersion $k(\omega)$ into a non-dispersive part $k_l(\omega)$ and a dispersive part $k_d(\omega)$:
\begin{equation}
    k(\omega, x) =
    \begin{cases}
      k_l(\omega), & \text{if } |x| \geq L, \\
      k_l(\omega) + k_d(\omega)(1-|x|/L), & \text{if } |x| \leq L.
    \end{cases}
    \label{eq:linear-ramp}
\end{equation}
The non-dispersive part $k_{l}(\omega) = \omega / v_\mathrm{g}$ determines the arrival time of the pulse; the dispersive part $k_{d}(\omega)$ is given by the nonlinear phase shift required to transform an exponentially increasing pulse into an exponentially decreasing one, which we have calculated in Eq.~(\ref{eq:dispersion}) to be $\Delta \phi(\omega) = -2 \arctan \!\big[(\omega - \wq)/\gamma\big]$. Integrating $k_{d}(\omega)(1 - |x|/L)$ over the region $x \in [-L, L]$ to match this phase, we obtain $k_{d}(\omega) = \Delta \phi(\omega)/L$. This dispersion relation ensures the right pulse transformation.

We seek a solution of wave propagating in the linear ramp of Eq.~(\ref{eq:linear-ramp}). The key equation is the wave equation 
\begin{equation}
    \frac{{d}^2}{{d}x^2} E(x) + k^2(x) E(x) = 0,
    \label{eq:wave-equation}
\end{equation}
where for simplicity we drop the $\omega$ argument and assume everything implicitly depends on $\omega$. The solution of Eq.~(\ref{eq:wave-equation}) can be analytically derived. For this, we follow Ref.~\cite{yeh1988optical}, which shows fields in a monotonic linear ramp (i.e., either $x\in[-L,0]$ or $x\in[0,L]$) can be expanded as $E(x) = b_1J(x) + b_2Y(x)$ where $J(x) = \sqrt{k(x)}\, J_{1/4}\!\left(\tfrac{k^{2}(x)}{2\alpha}\right)$, 
$Y(x) = \sqrt{k(x)}\, Y_{1/4}\!\left(\tfrac{k^{2}(x)}{2\alpha}\right)$ 
and $\alpha= k_d / L$. Here, $J_{1/4}(x)$ and $Y_{1/4}(x)$ are Bessel functions of the first and second kinds, respectively. From this, we can make the following ansatz for Eq.~(\ref{eq:wave-equation}): 
\begin{equation}
    E(x) =
    \begin{cases}
       b_5 e^{i k_{l} x} + b_6 e^{-i k_{l} x}, & x < -L, \\
       b_1 J(x) + b_2 Y(x), & -L < x < 0, \\
       b_3 J(x) + b_4 Y(x), & 0 < x < L, \\
       b_7 e^{i k_{l} x}, & L < x.
    \end{cases}    
\end{equation}
We solve the unknown coefficients by matching boundary conditions. We then calculate the reflection and transmission coefficients: $r = b_6e^{2ik_lL} / b_5$ and $t = b_7e^{2ik_lL} / b_5$. Their asymptotes at $L \gg \lambda = 2\pi v_\mathrm{g} / \omega$ are
\begin{align}
r(\omega) &= 
\frac{ i e^{i(k_l+k_{s})L} k_{d} \big[ k_l^{2} - k_{s}^{2} \cos\!\big(L(k_l+k_{s})\big) \big] }
     { 2L k_l^{2} k_{s}^{2} }, \\[6pt]
t(\omega) &= e^{i(k_l+k_{s})L}, \label{eq:tw}
\end{align}
where $k_s = k_l + k_d$. 
In this limit, the spatial variation of our dispersion $k(x)$ satisfies the adiabatic condition $|{d}k(x)/{d}x| \ll |k(x)|^2$, under which the transmission $t(\omega)$ is guaranteed to be unitary and  accumulates a phase that is proportional to the optical thickness~\cite{yeh1988optical}, which is also what we see in Eq.~(\ref{eq:tw}). As for the reflection, since $k_d = \Delta\phi(\omega) / L$, we can bound it by 
$    R(\omega) = |r(\omega)|^2 \leq \frac{[\Delta \phi(\omega)]^2}{L^4k_l^4}$,
which shows a $1/L^4$ dependence, guaranteeing near-unity transmission for long enough $L$. 
We plot $R(\omega)$ for three frequencies in Fig.~\ref{fig:3}(c) and show they all decay to zero in the limit of $L \gg \lambda_\mathrm{q}$.

We model the quantum dynamics using the Hamiltonian for qubits located outside the inhomogeneous region,
\begin{align*}
    {H} =\ &\wq\sum_{i=1}^2\sigma_i^+\sigma_i^-+\int_0^\infty d{k}\, v_\mathrm{g} k\,\left(a^\dagger_{\mathrm{R}k}a_{\mathrm{R}k}^{\phantom{\dagger}}+a^\dagger_{\mathrm{L}k}a^{\phantom{\dagger}}_{\mathrm{L}k}\right)\notag\\ &+g\int d{k}\left(e^{ikx_1}\sigma_1^\dagger a^{\phantom{\dagger}}_{\mathrm{R}k}+\mathrm{H.c.}\right)\notag\\
    &+g\int d{k}\left(b_7e^{ikx_2}\sigma_2^\dagger a^{\phantom{\dagger}}_{\mathrm{R}k}+b_6 e^{ikx_2}\sigma_2^\dagger a^{\phantom{\dagger}}_{\mathrm{L}k}+\mathrm{H.c.}\right).
\end{align*}
Here, $a_{\mathrm{R}k}^{\phantom{\dagger}}$ and $a_{\mathrm{L}k}^{\phantom{\dagger}}$ denote the right- and left-propagating modes without back-reflection. With inhomogeneity, however, the left-propagating eigenmode acquires a right-propagating component that couples to the second qubit.

\subsection{Sensitivity analysis}
\begin{figure}[t!]
    \centering
    \includegraphics[width=\linewidth]{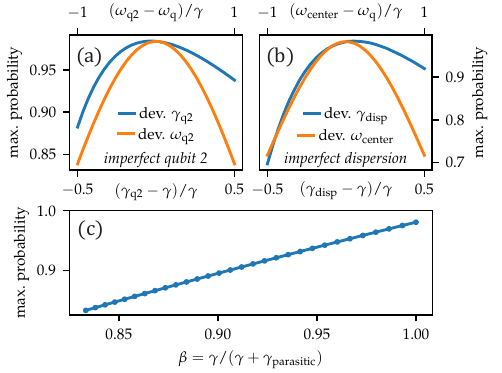}
    \caption{Maximal excitation probability when (a) the qubit 2 deviates in its transition frequency $\omega_\mathrm{q2}$ or coupling rate $\gamma_\mathrm{q2}$, (b) when the dispersion deviates in  its center frequency $\gamma_{\rm center} $ or linewidth $\gamma_{\rm disp}$, and (c) the qubits couple to the waveguide with loss characterized by a parasitic decay rate $\gamma_{\rm parasitic}$.}
    \label{fig:4}
\end{figure}

Besides the distance variation studied in the main text, other possible imperfections include deviations in the qubit parameters or the dispersion profile. 
To study these additional effects, we assume qubit separation $d = 5v_\mathrm{g}/\gamma$, coupling rate $\gamma=\pi\times10^{-4}\omega_\mathrm{q}$, and dispersion $\omega(k)$ in the form of Eq.~(\ref{eq:dispersion}). 
We detune the second qubit's coupling rate $\gamma_{q2}$ and transition frequency $\omega_{q2}$ from their supposed values of $\gamma$ and $\omega_\mathrm{q}$, and plot the corresponding maximal qubit excitation probability in Fig.~\ref{fig:4}(a). We also detune the dispersion's linewidth $\gamma_{\rm disp}$ and center frequency $\omega_{\rm center}$ from their supposed values of $\gamma$ and $\omega_\mathrm{q}$ in Eq.~(\ref{eq:dispersion}), and plot the corresponding maximal qubit excitation probability in Fig.~\ref{fig:4}(b). 
In both cases, we find our scheme is relatively robust: the transfer fidelity decreases at most 10\% even when the coupling rate or transition frequency of the qubit deviates by $0.5\gamma$, or the dispersion linewidth deviates by $0.25\gamma$, or its center frequency deviates by $0.5\gamma$.
We numerically confirmed that this 10\%-decrease is representative for $d$ from $v_g/\gamma$ to $9v_g/\gamma$, for all but the 0.25$\gamma$-deviation in the dispersion linewidth, whose fidelity-decrease worsens at short $d$ and reaches 20\% at $d = v_g/\gamma$.


\subsection{Loss analysis}
Our state-transfer scheme is mainly limited by two types of loss. The first is coupling loss during qubit emission, due to free-space radiation or imperfect chiral coupling. To quantify its effect, we use the same parameters as in Figs.~\ref{fig:4}(a) and \ref{fig:4}(b), but include for both qubits a parasitic decay channel $\gamma_{\rm parasitic}$ that does not couple to the target waveguide modes. 
Figure~\ref{fig:4}(c) shows that the simulated maximal excitation probability scales approximately linearly with the parameter $\beta = \gamma / (\gamma + \gamma_{\rm parasitic})$, corresponding to a coupling-induced loss of about 1\%--10\%  for realistic parasitic decay rates~\cite{mirhosseini_cavity_2019,scarpelli_99_2019}.

The second loss mechanism is photon loss in the waveguide, caused by scattering and absorption. This effect is expected to be particularly important in dispersion-engineered waveguides. Its impact depends on the specific implementation. Here we give a simple estimate for a waveguide formed by coupled resonators~\cite{mirhosseiniSuperconductingMetamaterialsWaveguide2018, poonTransmissionGroupDelay2006,jang_synchronization_2018,herrmann_arbitrary_2024}.

In such a system, the photon has a decay time $\tau = Q/\omega$, where $Q$ is the internal quality factor of each resonator and $\omega$ is the photon frequency. Using the relation derived after Eq.~(\ref{eq:dispersion}) in the main text, $\Delta t = d/v_\mathrm{g} + 2/\gamma$, and setting $\Delta t = \tau$, we estimate the maximal propagation distance before loss dominates as
\begin{equation}
    d_{\rm max} = (\gamma\tau - 2)v_\mathrm{g}/\gamma.
\end{equation}
Equivalently, the factor $\gamma\tau - 2$ gives the maximal distance in units of the pulse width $v_\mathrm{g}/\gamma$. For $d_{\rm max}>0$, one requires $\tau \ge 2/\gamma$, consistent with the fact that reversing the exponential pulse requires delaying its leading edge by at least twice its temporal width.

To estimate $d_{\rm max}$ in practice, we consider representative optical and microwave parameters. Resonators with quality factors as high as $Q\sim10^6$ have been reported in both domains~\cite{takesue_-chip_2013,zikiy_high-q_2023}, corresponding to $\tau\sim10\,\mathrm{ns}$ at optical frequencies and $\tau\sim1\,\mathrm{ms}$ at microwave frequencies. In the microwave domain, typical emission rates are $\gamma\sim10$--$100\,\mathrm{MHz}$~\cite{astafiev_resonance_2010,van_loo_photon-mediated_2013,mirhosseini_cavity_2019}, giving $d_{\rm max}\sim(10^4\text{--}10^5)\,v_\mathrm{g}/\gamma$, so the pulse can propagate over $10^4$--$10^5$ pulse widths before significant attenuation. In the optical domain, typical bare emission rates are only $\gamma\sim1$--$100\,\mathrm{MHz}$~\cite{hoang_ultrafast_2016}, which gives $d_{\rm max}<0$, indicating that even the shortest transfer would be strongly affected by propagation loss. With engineered optical emission, however, $\gamma$ can reach $\sim5$--$44\,\mathrm{GHz}$~\cite{zhangStronglyCavityEnhancedSpontaneous2018,liu_high_2018}, yielding $d_{\rm max}\sim(0.5\text{--}4.4)\times10^2\,v_\mathrm{g}/\gamma$, so the photon can still propagate over hundreds of pulse widths before significant loss.

\end{document}